\documentclass[pra,onecolumn,superscriptaddress,notitlepage,showpacs]{revtex4-1}
\usepackage{amsmath,mathrsfs,amsbsy,color,graphicx,bm,amsthm,amsfonts,dsfont}
\usepackage{times}
\usepackage{units}
\usepackage{braket}
\usepackage{amscd}
\usepackage{enumerate}

\begin{document}


\title{Entanglement Swapping with Independent Sources over an Optical Fibre Network}

\author{Qi-Chao Sun}
\affiliation{Department of Physics and Astronomy, Shanghai Jiao Tong University, Shanghai, 200240, China}
\affiliation{National Laboratory for Physical Sciences at Microscale and Department of Modern Physics, Shanghai Branch, University of Science and Technology of China, Hefei, Anhui 230026, China}
\affiliation{CAS Center for Excellence and Synergetic Innovation Center in Quantum Information and Quantum Physics, Shanghai Branch, University of Science and Technology of China, Hefei, Anhui 230026, China}

\author{Ya-Li Mao}
\affiliation{National Laboratory for Physical Sciences at Microscale and Department of Modern Physics, Shanghai Branch, University of Science and Technology of China, Hefei, Anhui 230026, China}
\affiliation{CAS Center for Excellence and Synergetic Innovation Center in Quantum Information and Quantum Physics, Shanghai Branch, University of Science and Technology of China, Hefei, Anhui 230026, China}

\author{Yang-Fan Jiang}
\affiliation{National Laboratory for Physical Sciences at Microscale and Department of Modern Physics, Shanghai Branch, University of Science and Technology of China, Hefei, Anhui 230026, China}
\affiliation{CAS Center for Excellence and Synergetic Innovation Center in Quantum Information and Quantum Physics, Shanghai Branch, University of Science and Technology of China, Hefei, Anhui 230026, China}

\author{Qi Zhao}
\affiliation{Center for Quantum Information, Institute for Interdisciplinary Information Sciences, Tsinghua University, Beijing, 100084, China}

\author{Sijing Chen}
\affiliation{State Key Laboratory of Functional Materials for Informatics, Shanghai Institute of Microsystem and Information Technology, Chinese Academy of Sciences, Shanghai 200050, China}

\author{Wei Zhang}
\affiliation{Tsinghua National Laboratory for Information Science and Technology, Department of Electronic Engineering, Tsinghua University, Beijing 100084, China}

\author{Weijun Zhang}
\affiliation{State Key Laboratory of Functional Materials for Informatics, Shanghai Institute of Microsystem and Information Technology, Chinese Academy of Sciences, Shanghai 200050, China}

\author{Xiao Jiang}
\affiliation{National Laboratory for Physical Sciences at Microscale and Department of Modern Physics, Shanghai Branch, University of Science and Technology of China, Hefei, Anhui 230026, China}
\affiliation{CAS Center for Excellence and Synergetic Innovation Center in Quantum Information and Quantum Physics, Shanghai Branch, University of Science and Technology of China, Hefei, Anhui 230026, China}

\author{Teng-Yun Chen}
\affiliation{National Laboratory for Physical Sciences at Microscale and Department of Modern Physics, Shanghai Branch, University of Science and Technology of China, Hefei, Anhui 230026, China}
\affiliation{CAS Center for Excellence and Synergetic Innovation Center in Quantum Information and Quantum Physics, Shanghai Branch, University of Science and Technology of China, Hefei, Anhui 230026, China}

\author{Lixing You}
\affiliation{State Key Laboratory of Functional Materials for Informatics, Shanghai Institute of Microsystem and Information Technology, Chinese Academy of Sciences, Shanghai 200050, China}

\author{Li Li}
\affiliation{National Laboratory for Physical Sciences at Microscale and Department of Modern Physics, Shanghai Branch, University of Science and Technology of China, Hefei, Anhui 230026, China}
\affiliation{CAS Center for Excellence and Synergetic Innovation Center in Quantum Information and Quantum Physics, Shanghai Branch, University of Science and Technology of China, Hefei, Anhui 230026, China}

\author{Yidong Huang}
\affiliation{Tsinghua National Laboratory for Information Science and Technology, Department of Electronic Engineering, Tsinghua University, Beijing 100084, China}

\author{Xianfeng Chen}
\affiliation{Department of Physics and Astronomy, Shanghai Jiao Tong University, Shanghai, 200240, China}

\author{Zhen Wang}
\affiliation{State Key Laboratory of Functional Materials for Informatics, Shanghai Institute of Microsystem and Information Technology, Chinese Academy of Sciences, Shanghai 200050, China}

\author{Xiongfeng Ma}
\affiliation{Center for Quantum Information, Institute for Interdisciplinary Information Sciences, Tsinghua University, Beijing, 100084, China}

\author{Qiang Zhang}
\affiliation{National Laboratory for Physical Sciences at Microscale and Department of Modern Physics, Shanghai Branch, University of Science and Technology of China, Hefei, Anhui 230026, China}
\affiliation{CAS Center for Excellence and Synergetic Innovation Center in Quantum Information and Quantum Physics, Shanghai Branch, University of Science and Technology of China, Hefei, Anhui 230026, China}

\author{Jian-Wei Pan}
\affiliation{National Laboratory for Physical Sciences at Microscale and Department of Modern Physics, Shanghai Branch, University of Science and Technology of China, Hefei, Anhui 230026, China}
\affiliation{CAS Center for Excellence and Synergetic Innovation Center in Quantum Information and Quantum Physics, Shanghai Branch, University of Science and Technology of China, Hefei, Anhui 230026, China}

\date{\today}

\begin{abstract}
Teleportation of an entangled state, known as  entanglement swapping, plays an essential role in quantum communication and network. Here we report a field-test entanglement swapping experiment with two independent telecommunication band entangled photon-pair sources over the optical  fibre network of  Hefei city. The two sources are located at two  nodes 12 km apart  and the Bell-state measurement is performed in a third location  which is connected to the two source nodes with 14.7 km and 10.6 km optical fibres. An average visibility of $79.9\pm4.8\%$ is observed in our experiment, which is high enough to infer a violation of Bell inequality. With the entanglement swapping setup, we demonstrate a source independent quantum key distribution, which is also immune to any attack against detection in the  measurement site.
\end{abstract}

\maketitle

With the help of quantum entanglement and Bell state measurement (BSM), entanglement swapping~\cite{Bennett93Teleportation,ZukowskiEntanglementSwapping} can entangle two remote particles sharing no common history without interacting them. Therefore, except for its interest in the fundamental study of quantum information science, entanglement swapping also constitutes a core part for quantum repeater~\cite{Briegel1998QuantumRepeater,DLCZ} in long distance quantum communication~\cite{Gisin2007QuantumCom} and quantum internet~\cite{Kimble2008QuanInt} for distributed quantum computation~\cite{Nielsen2010QCQIBook}.

For all these quantum communication applications, photons are the most suitable carriers due to its flying nature and robustness against environmental decoherence. In the past two decades, various photonic entanglement swapping experiments have been implemented~\cite{Pan1998EntanglementSwap,Riedmatten2005Swapping,Yang2006IndSource,Kaltenbaek2009HFEntSwap,Halder2007SwappingCWPump}. Recently, a field test of entanglement swapping was demonstrated through free space link~\cite{Herbst2015EntSwap143}. However, in most of the previous demonstrations of entanglement swapping, the same  femtosecond laser was used in producing two entangled photon-pair sources~\cite{Pan1998EntanglementSwap,Riedmatten2005Swapping}, making them impractical to be implemented in quantum network where the sources should be placed in distant nodes. Pioneering works have been done to synchronize two sources which are pumped independently in laboratory~\cite{Yang2006IndSource,Kaltenbaek2006IndSource,Halder2007SwappingCWPump,Kaltenbaek2009HFEntSwap}. Owing to the difficulties in guaranteeing the indistinguishability of photons after they are transmitted through the realistic channel, a field test of entanglement swapping with independent entangled photon-pair sources remains an experimental challenge.

Besides its applications in quantum repeater, the nonlocal correlation created by entanglement swapping can also be employed in quantum cryptography. Although quantum key distribution (QKD) can in principle provide information-theoretical security based on quantum mechanics, practical QKD systems are suffering from side-channel attacks that explore the device flaws in real-life implementation. Recently, the detector side-channel attacks have been removed by employing  measurement-device-independent (MDI) QKD protocol~\cite{lo2012MDI}. However, potential loopholes still exist on the source side. When the photon source is not well prepared according to the security proof model, attacks would be possible~\cite{Tang2013Source}.

When entanglement swapping is used for QKD, each of Alice and Bob employs an entangled photon-pair source and sends one of the twin photons to an untrusted measurement site, Eve. Similar to the scenario of MDIQKD, the security of the final key does not rely on how faithful Eve performs the BSM or announces the results. The MDI property stems from the beauty of teleportation, in which the measurement site does not reveal any information about the transmitted state in order to ensure the quantum transmission is faithful. Furthermore, Alice and Bob measure the rest photon locally to obtain raw data which enjoys the source-independent property. In fact, an entangled source can be used as a basis-independent source~\cite{koashi2003secure,ma2007quantum}. From this point of view, the entanglement swapping setting for QKD enjoys both MDI and source-independent security properties.

%



\begin{figure}[!htbp]
\centering \includegraphics[width=0.45\textwidth]{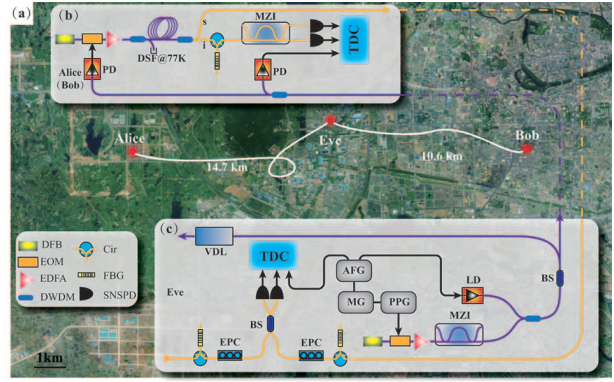}
\caption{Bird's-eye view and schematics of entanglement swapping in Hefei optical fibre network. (a) Experimental scheme. The two end-users, Alice and Bob, are located in Hefei Innovation Industrial Park  (N~$31^{\circ}50^{'}11.42^{''}$, E~$117^{\circ}7^{'}54.37^{''}$) and University of Science and Technology of China (N~$31^{\circ}50^{'}7.50^{''}$, E~$117^{\circ}15^{'}50.56^{''}$), respectively. They prepare entangled photon-pairs, detect the idler photons (i) locally and send the signal photons (s) to Eve, located in Hefei Software Park (N~$31^{\circ}51^{'}$ $5.42^{''}$, E~$117^{\circ}11^{'}55.82^{''}$), to perform the BSM. (b) and (c) Experimental setups of time-bin entangled photon-pair source and BSM, respectively. The quantum and classical channels are represented by yellow and purple lines, respectively. The arbitrary function generator (AFG) synchronized by the microwave generator generates 10 MHz clocks for the  measurement devices, which is distributed to Alice and Bob through the same optical fibre with  driving signals of sources using DWDM technique.}
\label{fig:setup}
\end{figure}

In this Letter, we present the realization of entanglement swapping with two independent telecommunication band entangled photon-pair sources placed at two nodes 12 km apart in Hefei optical fibre network (marked as Alice and Bob in Fig.~\ref{fig:setup}(a)). A third node, Eve, in-between of Alice and Bob performs the BSM. The idler photons of entangled photon-pairs of Alice and Bob are detected locally and the signal photons are sent to Eve over optical fibre with length of 14.7 km and 10.6 km, respectively.

Both Alice and Bob generate time-bin entangled photon-pair through spontaneous four-wave-mixing (SFWM) in a 300 m dispersion shifted fibre (DSF) pumped by two consecutive laser pulses as shown in Fig.\ref{fig:setup}(b). The two consecutive laser pulses are generated  by carving a CW laser beam ($\lambda=1552.54$ nm) emitted by a distributed feedback laser (DFB) using an electro-optical modulator (EOM) and then amplified by an Erbium doped fibre amplifier (EDFA). The amplified spontaneous emission noise photons of EDFA are suppressed by using cascaded dense wavelength division-multiplexing (DWDM) filters. Because the phase matching condition of SFWM is eased owing to zero dispersion of DSF in telecommunication band, the down converted photon-pair has a broad bandwidth. In addition, the generated photon-pairs are accompanied by  phonon-related single photons.  Another set of cascaded DWDM filters is used to single out photons with wavelength of $1555.73\ \mathrm{nm}$ (idler) and $1549.36\ \mathrm{nm}$ (signal), as well as reject the other photons with extinction ratio of up to 115 dB. Furthermore, the DSF is immersed in liquid nitrogen for further reduction of phonon-related photons.  Since the time delay between the two consecutive laser pulses is 1 ns, much less than the coherence time of the CW laser, their relative phase is well-defined ($\delta=0$). The down converted photon-pairs originated from the two laser pulses are in superposition state $\ket{\phi^+}_{s,i}=\frac{1}{\sqrt{2}}(\ket{1,1}_{s,i}+\ket{2,2}_{s,i})$, where $\ket{1,1}_{s,i}$ and $\ket{2,2}_{s,i}$ represent the state of  photon-pair created by the first pulse and the second pulse of the two consecutive pulses, respectively.

Eve performs the BSM by interfering the signal photons at a 50:50 beam splitter (BS) and detecting them with two superconducting nanowire single photon detectors (SNSPD), as shown in Fig.~\ref{fig:setup}(c).  This scheme can, in principle, discriminate two Bell states, $\ket{\Psi^{\pm}}=\frac{1}{\sqrt{2}}(\ket{1,2}\pm\ket{2,1})$, out of the four Bell states (the other two Bell states are defined by $\ket{\Phi^{\pm}}=\frac{1}{\sqrt{2}}(\ket{1,1}\pm\ket{2,2})$ ). Here, we measure the time of detection events using a time-to-digital converter (TDC) with a resolution of 4 ps and only select the projections onto the singlet state $\ket{\Psi^-}$. Consequently, the two corresponding idler photons are projected onto Bell state $\ket{\Psi^+}$.

Achieving two-photon interference with high and stable visibility in BSM is critical for entanglement swapping experiment.  This demands to eliminate distinguishability of the two photons in every degree of freedom. In our experiment, the spatial and spectrum indistinguishablilities are guaranteed by using single mode fibre as spatial filter and identical fibre Bragg gratings (FBG) as spectrum filters, respectively.  However, the real challenge is to eliminate the distinguishability between independent photons in the time domain at interference. The temporal distinguishability arises from the sharp time correlation between the twin photons  is suppressed by filtering the photons with  FBGs with bandwidth of 4 GHz for both signal and idler photons. The bandwidth of FBGs is half of that of the pump pulse, meaning the single-photon-state purity is about $99.4\%$. To synchronize the two sources, we use a microwave generator (MG) to provide the master clock for a pulse pattern generator (PPG)  in the node of Eve (see Fig.~\ref{fig:setup}(c)), which drives an EOM to carve a CW laser beam into 300 MHz laser pulse train with pulse width of about 75 ps. After passing through a Mach-Zehnder interferometer (MZI) with 1 ns path difference, each laser pulse is split into two consecutive laser pulses. Then the laser pulse train is sent to Alice and Bob through optical fibre and  detected by using 45 GHz photo-detectors (PD) to generate driving signals for the entangled photon-pair sources.  Compared to previous experiments to synchronize photon sources which were hard to maintain even in the laboratory~\cite{Yang2006IndSource,Kaltenbaek2006IndSource,Halder2007SwappingCWPump,Kaltenbaek2009HFEntSwap}, in our field test, photons are interfered after being transmitted through optical fibres from Alice to Eve and from Bob to Eve which are separated by 6.4 km and 6.5 km in the field, respectively. The surrounding temperature variance and vibration give rise to dramatical fluctuation in the effective length and polarization property of the optical fibre. We feed back the error signal derived from photon arrival time to the variable delay line (VDL)  to control the creation of Bob's  entangled photon-pair source for every 200 s. The fluctuation of polarization is reduced using electronic controlled polarization controllers (EPC) in a separate feedback loop. With these feedbacks, we largely suppress the distinguishability. In addition, automation is also applied to stabilize the phase of MZI, the power stability of pump power, the central wavelength of FBG, and to optimize the extinction ratio of EOM.

\begin{figure}
\centering
\includegraphics[width=0.45\textwidth]{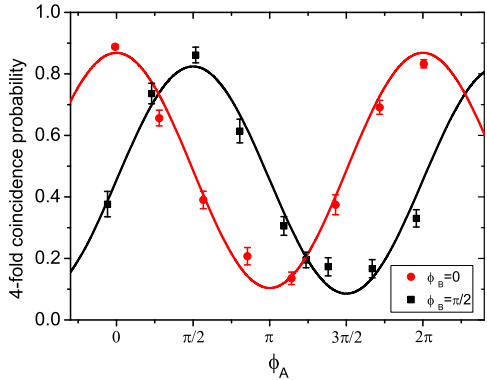}
\caption{Four-fold coincidence count probabilities as a function of phase of Alice's  MZI. The error bars indicate
one standard deviation  calculated from measured counts assuming Poissonian detection statistics. Each data point is accumulated for more than 10 hours. The visibilities of the fitted curve are $81.2\pm6.2\%$ and $78.6\pm7.3\%$ for measured results with $\phi_B=0$ and $\phi_B=\pi/2$, respectively. }
\label{fig:IntCur}
\end{figure}

To verify successful entanglement swapping, we measure the correlation curve of the two idler photons conditioned on the projection of the two signal photons onto Bell state $\ket{\Psi^-}$. In this stage, Bob fixes the phase of his MZI and Alice sweeps the phase of her MZI. The MZIs used in our experiment are plannar lightwave circuit devices, and the phase can be tuned by controlling the device temperature. The detection signals are measured by using a TDC with 4 ps time resolution. We implement a four-fold coincidence measurement based on measurement results of Alice, Bob and Eve, and the results are depicted in Fig.~\ref{fig:IntCur}.  The probabilities of four-fold coincidence count of  photons  show sinusoidal curves with fitted visibilities of $81.2\pm6.2\%$ and $78.6\pm7.3\%$ for measurement results when the phase of Bob's MZI is set as $0$ and $\pi/2$, respectively. The average visibility of the fitted curves is $79.9\pm4.8\%$, which corresponds to a fidelity of $84.9\pm3.6\%$ ($F=(3V+1)/4$). This value exceeds the classical limit of $1/3$ to demonstrate entanglement~\cite{Peres1996Separability} and infers a violation of CHSH Bell inequality by more than two standard deviations, provided the swapped photons are in Werner state~\cite{CHSH,Marcikic2004EntDis}. In our experiment, the visibility of the correlation fringe is mainly degraded by multi-photon events. The average photon-pair number per pulse is about 0.03 for both entangled photon-pair sources, which upper-bounds the visibility of two-photon interference to be 0.91 in our experiment.

In the second stage of our experiment, we demonstrate QKD with swapped photons. Actually, in our experiment the entanglment between the idler photons is generated a posteriori~\cite{Peres2000EntSwapDelCho}. By measuring the idler photons in the time basis, $\lbrace\ket{1},\ket{2}\rbrace$ and energy basis, $\{(\ket{1}\pm\ket{2})/\sqrt{2}\}$, Alice and Bob extract a secure key from their local measurement results conditioned on the BSM outcomes.  As mentioned in Ref.~\cite{ma2007quantum}, the entangled photon-pair source can offer basis-independent source for QKD. Thus, following the security proof from Koashi and Preskill \cite{koashi2003secure}, the final key rate is given by,
\begin{equation}
\begin{aligned}
R\ge Q[1-f H(e_b)-H(e_p)],
\end{aligned}
\end{equation}
where $Q$ is the sifted key rate, $f$ is the error correction efficiency (we use $f=1.16$ here), $e_b$ and $e_p$ are the bit and phase error rates, respectively, and $H(x)=-x\log_2(x)-(1-x)\log_2(1-x)$ is the binary entropy function. The estimation of phase error rate is the key to the security analysis \cite{shor2000simple}. Due to the symmetry between the complementary bases, the phase error rate of one basis can be estimated by the bit error rate of the other basis. The details of the analysis is shown in Appendix \ref{serfling}.

\begin{figure}[!htbp]
\centering
\includegraphics[width=0.45\textwidth]{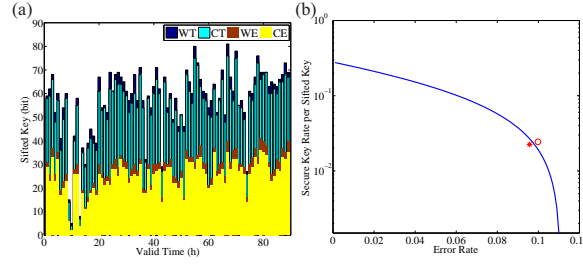}
\caption{Experimental results. (a) Sifted key in the time basis (T) and energy basis (E) per hour, with the numbers of wrong (W) and correct (C) bits. (b) Secure key rate per sifted key with one B step of the entanglement swapping based QKD as well as the simulation result. The circle and star correspond to the experimental result in energy basis (60 bits) and the time basis (58 bits), respectively. In the simulation, the parameters are set as $N^e=N^t=2500$ and $e_b^e=e_b^t$.}
\label{fig:keyrate}
\end{figure}

The experimental data is accumulated for 89 hours, and the results are shown in Table \ref{tab:Rawkey} and Fig.~\ref{fig:keyrate}. The total number of sifted key is 5096 bits. The error rate in energy basis is consistent with the average visibility of correlation fringe according to the relation  $e_b=(1-V)/2$. Due to the slightly difference between the error rates of sifted key in the two bases, the secure key rates are evaluated for the two bases separately. In postprocessing, we apply the Gottesman-Lo security analysis method using two-way classical communication \cite{gottesman2003proof}. The details are shown in Appendix~\ref{bstep}. As shown in Fig.~\ref{fig:keyrate}(b), 118 bits of secure key are distilled from the sifted key.

\begin{table}[!htbp]
\caption{\label{tab:Rawkey}The number of sifted key bits (N) and error rate ($e_b$).  The superscripts $e$, $t$ and $tot$ denote values in the energy basis, time basis, and both of them together, respectively.}
\begin{ruledtabular}
\begin{tabular}{cccccc}
\textrm{$N^{tot}$}&\textrm{$N^e$}&\textrm{$N^t$}&\textrm{$e_b^e$}&\textrm{$e_b^t$}&\textrm{$e_b^{tot}$}\\
\colrule
5096&2485&2611&0.09980&0.09575&0.09772
\end{tabular}
\end{ruledtabular}
\end{table}

Here, we need to point out that in the security of such scheme is not fully device-independent. The underlying assumption is that Alice's and Bob's local detection efficiencies are independent of the basis choices \cite{ma2008quantum}. This assumption can be guaranteed when the entangled photon-pair source is single-mode or the local measurement device follows the squashing model~\cite{Beaudry2008Squashing}. In order to achieve fully device-independent QKD, high detection efficiency is required. In fact, a very recent loophole-free Bell inequality testing experiment has successfully coupled nitrogen-vacancy center quantum memory with 637 nm photon and generated entanglement between two electron spins 1.3 km apart~\cite{Hensen2015LoopholeFreeBT}. Moreover, in order to achieve a quantum repeater, entanglement swapping must combine with quantum memories~\cite{Briegel1998QuantumRepeater,DLCZ}. Note that our setup can be directly coupled with Erbium-doped quantum memory~\cite{Saglamyurek2015MemoryTelc}.

\section*{Acknowledgments}
The authors would like to thank H. Lu, X.-L. Wang, Q. Zhou and H.-L. Yin for enlightening discussions, especially to Y. Liu for his useful discussions and comments on the key rate analysis. This work was supported by the National Fundamental Research Program (under Grant No. 2013CB336800 and 2011CBA00303), the National Natural Science Foundation of China, the Chinese Academy of Science, the 10000-Plan of Shandong Province and Quantum Ctek Co., Ltd.

\appendix

\section{Phase error rate estimation} \label{serfling}
 In postprocessing, the phase error rate of one basis can be estimated by the bit error rate of the other basis. Here, we take the time basis for example to show how to estimate phase error rate. The analysis for the energy basis is the same.
Due to the symmetry between the time and energy bases, in the large data size limit, the phase error rate in the time basis, $e_p^t$, equals to the bit error rate of the energy basis, $e_b^e$,
\begin{equation}
\begin{aligned}
e_p^t=e_b^e=E
\end{aligned}
\end{equation}
where $E$ is the QBER in the energy basis. Considering the statistical fluctuation, there exists a gap between $e_b^e$ and $e_p^t$. Random sampling method provides a upper bound for this gap with a fixed failure probability $\epsilon$ \cite{ma2012statistical,curty2014finite}, in our security analysis, $\epsilon=10^{-10}$. Here, the Serfling inequality \cite{serfling1974probability} is applied to estimate this gap. The upper bound for $e_p^t$ is:
\begin{equation}
\begin{aligned}
e_p^t\le e_b^e+g(\epsilon,n_e,n_t)
\end{aligned}
\end{equation}
where $g(\epsilon,n_e,n_t)$ is the function of failure probability $\epsilon$, sample size $n_e$ and the other basis population size $n_t$.

Considering a finite list of values $x_1,\cdots , x_N$, for any $i$, $x_i\in [a,b]$, $n$ is the sample size and $N$ is the population. $X_1,\cdots, X_n$ are the values of chosen sample, $S_n$ is the summation of them, $S_n=\sum X_i$, $\mu=\frac{\sum_1^N x_i}{N}$ is the total average value and $f_n^*=n-1/N$ is the sampling fraction. For $k>0$, we have
\begin{equation}
\begin{aligned}
P_n(k)&=P(|S_n-n\mu|\ge nk),\\
P_n(k)&\le exp~[\frac{-2nk^2}{1-f_n^*}(b-a)].
\end{aligned}
\end{equation}

In our case, $e_{p}^t=\mu$, $e_{b}^e=S_n/n$, $a=0$, $b=1$, $n_e$ and $n_t$ are the numbers of raw key for energy basis and time basis, respectively, $N=n_e+n_t$ is the total raw key, thus
\begin{equation}
\begin{aligned}
Pr(e_{p}^t \ge e_{b}^e +k)\le exp~[- k^2 ~\frac{2n^e(n^e+n^t)}{n^t+1}].\\
\end{aligned}
\end{equation}
Consequently, the upper bound for $e_{p}$ is :
\begin{equation}
\begin{aligned}
e_{p}^t\le e_{b}^e+\sqrt{\frac{(n^t+1)\log(1/\epsilon)}{2n^e(n^e+n^t)}},
\end{aligned}
\end{equation}
and \begin{equation}
\begin{aligned}
g(\epsilon,n_e,n_t)=\sqrt{\frac{(n^t+1)\log(1/\epsilon)}{2n^e(n^e+n^t)}}.
\end{aligned}
\end{equation}

\section{B-step two way classical communication}\label{bstep}

In Shor and Preskill's \cite{shor2000simple} security proof, the maximal tolerable error rate is only 11\%  with one way classical communication (1-LOCC). In some practical cases, the bit error rate may be higher, or close to 11\%. It is too high to generate keys. Gottesman and Lo's security proof \cite{gottesman2003proof} shows that QKD with two-way classical communication (2-LOCC) can tolerate a much higher bit error rate than that with one-way classical communication (1-LOCC). Thus we apply the B step method to perform two-way classical communication and the final key rate after one B step is:
\begin{equation}
\label{eq:KeyRate}
\begin{aligned}
&R\ge \frac{p_s}{2} Q[1-f H(e_b')-H(e_p')]
\end{aligned}
\end{equation}
where $e_b^{'}$ and $e_p^{'}$ are the bit error rate and phase error rate after a B step, respectively and $p_s={e_b}^2+(1-e_b)^2$.

Classically, a B step involves random pairing of the key bits. The strings $x_1$, $x_2$ are on Alice'side and  $y_1$, $y_2$ are on Bob'side. Both Alice and Bob announce the parities, $x_1\oplus x_2$, $y_1\oplus y_2$. If their parities are the same, they keep $x_1$, $y_1$, otherwise they discard all of them. Note that at least half of the raw keys are discarded. After a B step, the bit error rate $e_b^{'}$ and the upper bound for phase error rate $e_p^{'}$ \cite{ma2006decoy} becomes:
\begin{equation}
\begin{aligned}
e_b^{'}&=\frac{{e_b}^2}{p_s},\\
e_p^{'}&\le 2\frac{e_p(1-e_p-e_b)}{p_s}.
\end{aligned}
\end{equation}\\

\bibliographystyle{apsrev4-1}
%

\end{document}